# Hacking the Sky

Robert J. Simpson

Department of Physics and Astronomy, Cardiff University

**Summary**. In this article I present some special astronomical scripts created for Google Earth, Google Sky and Twitter. These 'hacks' are examples of the ways in which such tools can be used either alone, in on conjunction with online services. The result of a combination of multiple, online services to form a new facility is called a mash-up. Some of what follows falls into that definition. As we move into an era of online data and tools, it is the network as a whole that becomes important. Tools emerging from this network can be capable of more than the sum of their parts.

## 1 Introduction

Applications with online links, like Google Earth[1], and online services such as Twitter[2], are a target for a growing online community of programmers. At one time, these people would have been considered hackers; coders who figure out the inner workings of a thing in order to change and modify it as desired. In the online age companies like Google and Twitter release application programming interfaces (APIs), which enable anyone to interact with their service. This enables them to create interesting or useful services based upon it.

---

[1] http://earth.google.com at time of press.
[2] http://www.Twitter.com at time of press.

Whilst many may not entirely obey the rules surrounding the use of these APIs, the community as a whole has created a plethora of diverting and remarkable services.

In this article, I will describe a sample of the kind of things I have personally created. There follows a description of the motivations, techniques and results behind several projects I have worked on at some point over the past year or two.

## 2 Live Satellite Tracking with Google Earth

The Google Earth application was originally known as Earth Viewer and was created by Keyhole Inc. Google purchased the company in 2004 and Google Earth was released in 2006 in its current form.

Since that time, Google Earth has become a widely used tool at home, in education and in industry. Google Earth allows users to create geocoded[3] XML data files (KML files) that enable the display of complex information or whole datasets in the Google Earth environment.

Google Earth's 3D projection allows for the accurate placement of objects in 3D, with a specific latitude, longitude and altitude. It also allows displayed data to be updated periodically, on timescales as frequent as 1 second.

### 2.1 Satellite Tracking Scripts

With this in mind, I created a satellite tracking service for Google Earth, hosted via my own website Orbiting Frog[4]. Initially this was an online script that accessed a service called CelesTrak[5]. Operated by the Center for Space Standards and Innovation (CSSI), CelesTrak provides up-to-date two-line element (TLE) datasets for all trackable satellites orbiting the Earth.

---

[3] Geocoding is the term for embedding geographical data in a web-accessible format.
[4] http://www.orbitingfrog.com at time of press.
[5] http://www.celestrak.com/ at time of press.

TLE datasets are small files which mathematically describe the position and motion of orbiting bodies. CelesTrak updates these daily and so acts as a reference for accurate satellite position data.

Originally my tracking script used the standard orbital elements model to decode the TLE data and output a Google Earth file describing the position of a satellite, e.g. the International Space Station. This was then updated once a minute and replotted on Google Earth to provide a live location for the space station on the virtual globe. Features such as flight path and viewing horizon were also added at this stage.

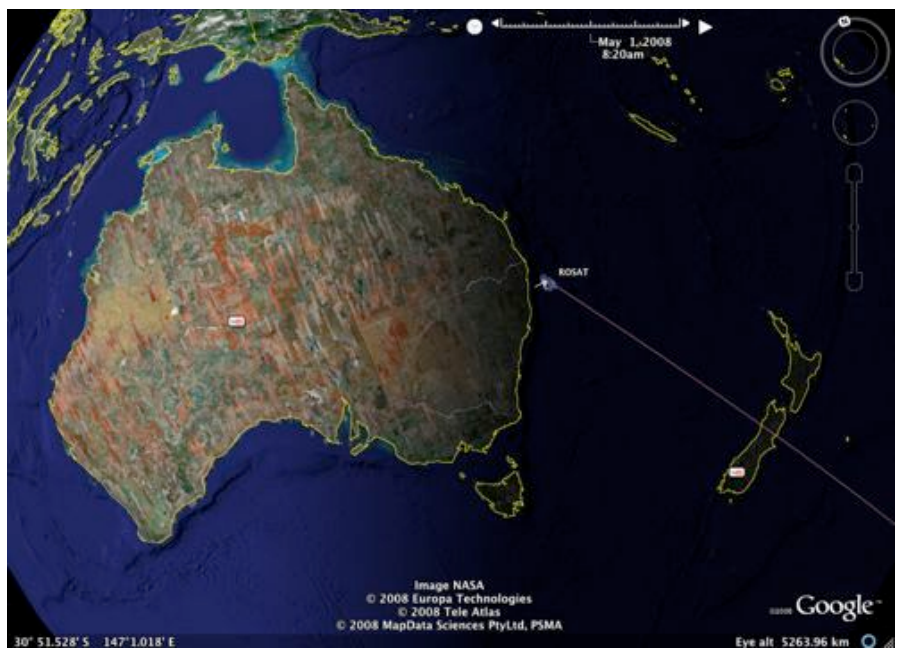

**Figure 1** – ROSAT orbiting high above Australia, as displayed in Google Earth.

It was then trivial to enable users to tell the script to access alternative TLE files, and thus the tool allowed any object to be tracked via Google Earth.

This led to some problems, mainly that the script became popular and began to frequently crash my server. One user in Western France was using it to track 10,000 objects in Google Earth on a daily basis!

I have since limited the general version of the script to parse only 100 objects at a time. This seems to have solved any problems.[6]

### 2.2 Interesting Objects

I created several pre-built tracking packages, which are available to download via Orbiting Frog. One package allows users to simply to track the International Space Station. Another tracks several notable science satellites and telescopes such as XMM-Newton, IRAS and ROSAT.

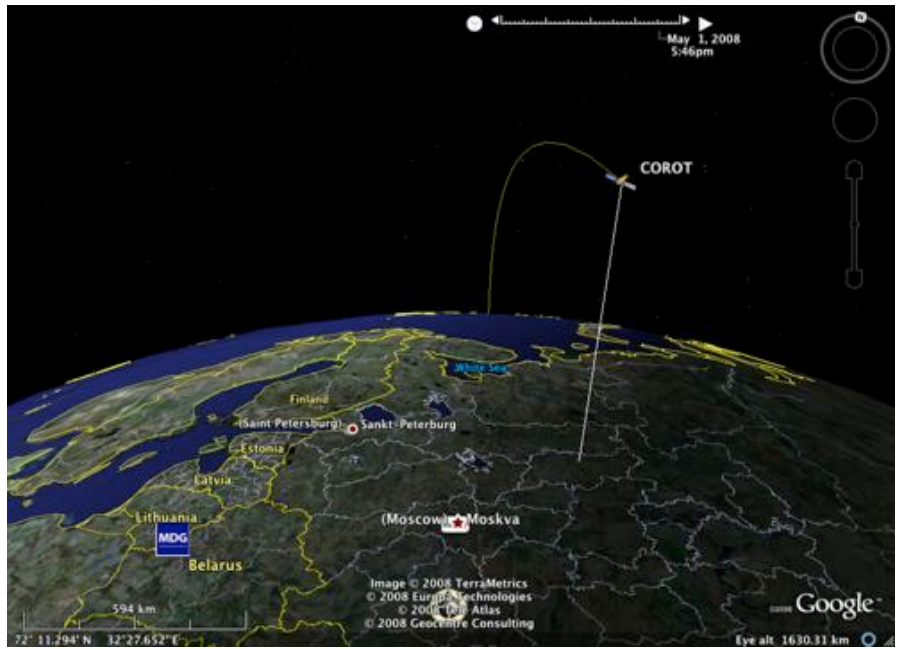

**Figure 2** – CoRoT passes over Russia, as displayed in Google Earth.

---

[6] The satellite tracking script could be found online at time of press at http://orbitingfrog.com/blog/satellite-kml/

One particularly popular implementation of the script is one that tracks the debris of the Chinese satellite Fengyun-1C. This satellite was destroyed in January 2007 when the Chinese government controversially tested a surface-to-space missile. The resultant debris were scattered in high-orbits and so much of it is still travelling above our heads years later.[7]

The three-dimensional nature of Google Earth enables people to begin to grasp the true scale of the Earth. Satellites, which we know are very high above us, appear to skim the surface of the virtual globe. When updating at a frequent interval this script also lets users see how quickly objects can orbit the planet and in what manner they do so.

I have used the tool in several schools to demonstrate many of the aspects of the human exploration of space. It is particularly effective when used in conjunction with a visible pass of the International Space Station overhead.

## 3 Google Sky Tools

In August 2007, Google Earth added a new feature, Google Sky (Scranton et al., 2007). This 'Sky' mode effectively places the user inside the virtual globe model and then projects the stars onto it, instead of the Earth.

By this time, the KML file system was fairly well understood by the community of users and it didn't take long for a series of novel applications of Google Sky to appear. Google worked with many of the early content creators and incorporated some of their scripts into future versions of the software. The current version of Google Earth (4.3 at time of press) includes a much more fully-featured Google Sky.

---

[7] The Chinese satellite debris tracker was found online at time of press at http://orbitingfrog.com/blog/2008/04/21/china-satellite-debris-in-google-earth/

### *3.1 VOEvents*

One interesting service that can be accessed from within Google Sky is the VOEvent network from eStar. Users can see live events pop up on the sky, such as Gamma Ray Bursts or alerts from the Optical Gravitational Lensing Experiment (OGLE). These pop-ups then allow the user to access further information and learn more about the science behind them.

### 3.2 SIMBAD

The SIMBAD service[8] has also been implemented within Google Sky. A simple and small file can be downloaded which allows you to perform cone searches for objects within the current Google Sky window. Each of these objects can then be explored further via a link to the SIMBAD website.

A similar idea would allow the creation of Google Sky services that could report back academic papers within a region, or indeed any other look-up involving geocoded (or in fact astrometry-coded) data.

### 3.3 Exploring Wavelengths

Originally one complaint made of Google Sky was its limitation to just optical wavelengths. So much of modern astronomy is concerned with the study of other parts of the electromagnetic spectrum that this omission by the Google Sky team was much-commented upon.

Google Sky rectified the problem by adding additional 'layers' for well-known infrared, microwave and x-ray frequencies. These could be switched on and then viewed on top of the optical data with a customized degree of transparency to allow for comparisons.

This feature is still in its early stages and will hopefully develop with future revisions of Google Sky.

---

[8] http://simbad.u-strasbg.fr/simbad/

## 4 SCUBA Data on Google Sky

In 2006, when I began my PhD, I attended a workshop at the Royal Observatory in Edinburgh, which reviewed the history of the SCUBA instrument on the JCMT. It also looked forward to SCUBA-2.

During this meeting I learned informally that all of the SCUBA data ever taken could fit onto a standard iPod (approximately 80GB). Later on, in 2007 I learned that Di Francesco et al. (2008) has re-reduced the entire SCUBA data catalogue in a consistent manner, creating a standardised atlas of the entire dataset.

Knowing that the entire set was reasonably small and that it had now been entirely reduced in a standard fashion, I realised that it would be a straightforward task to placed this new, complete catalogue onto Google Sky.

Any dataset that can be easily segmented or montaged, can be placed into a KML file format fairly directly. This is what I set about doing and it was complete in late 2007. The project was presented as a poster at the UK National Astronomical Meeting in 2008[9].

---

[9] It was to be found online at http://orbitingfrog.com/scuba at time of press

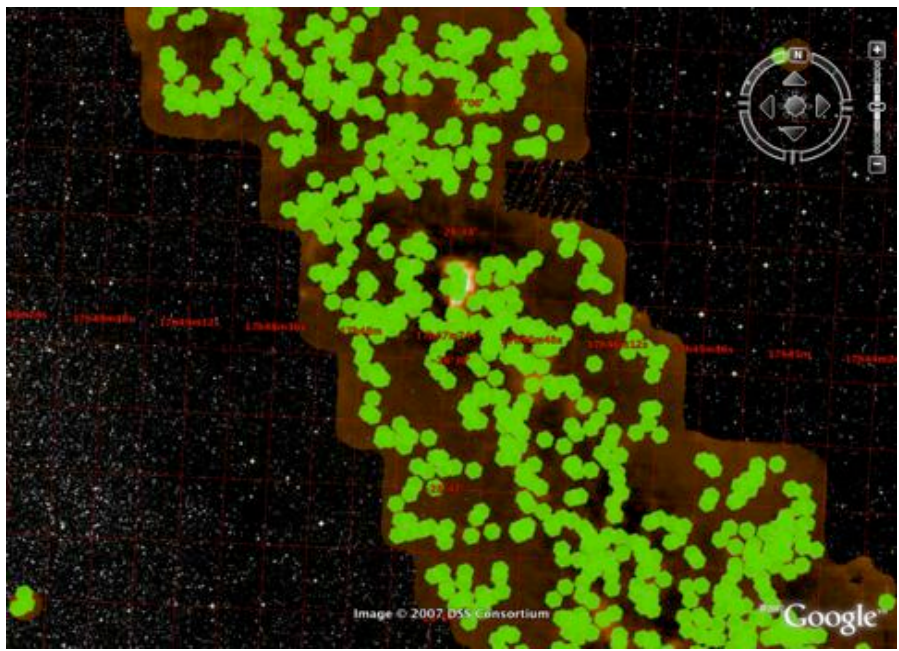

**Figure 3** – SCUBA Data on Google Sky for the area around the constellation Sagittarius. Green hexagons denote SCUBA point sources; the brown region shows the 850-micron SCUBA map of the region.

### 4.1 One Catalogue, Many Maps

Di Francesco (2008) produced a large catalogue with associated maps. Since SCUBA covered only a small fraction of the sky, the catalogue cannot be as intuitively browsed as many all-sky atlases. To get around this problem, the SCUBA Data on Google Sky is presented in two parts:

A downloaded catalogue of point-sources which sits on the local machine,

Dynamically loaded maps, which are only downloaded when the region of the sky is seen in Google Sky.

This gives the user the whole catalogue (where SCUBA sources are presented as small, green hexagons on the sky) and when inspected more closely, fully reduced 450- and 850-micron maps.

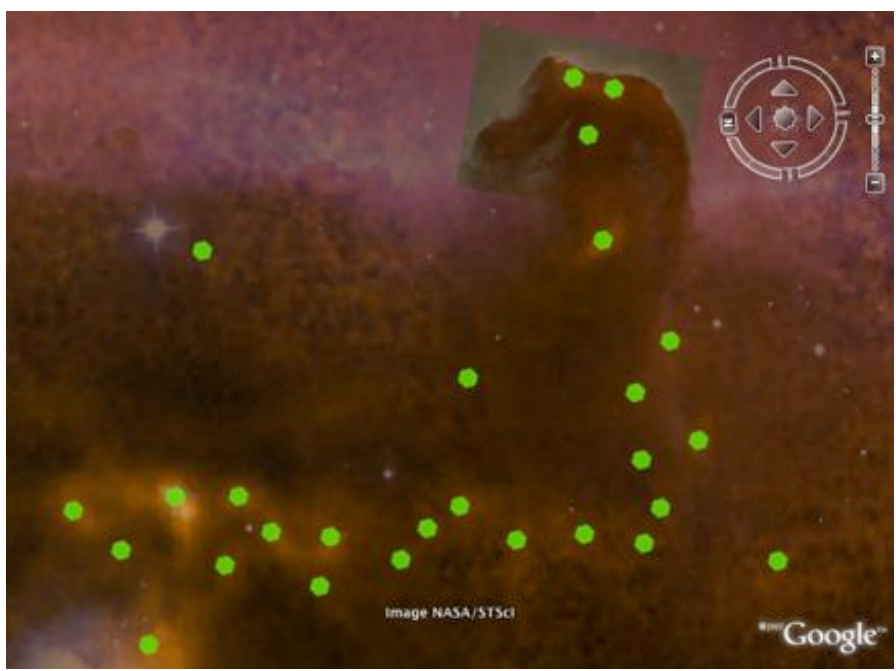

**Figure 4** – Data on Google Sky showing the Horsehead Nebula in Orion. Orange-brown SCUBA 850-micron data is overlaid on top of purple optical data. SCUBA shows the dust which is blocking out optical light beneath the Horse's head.

Also included in the downloadable file is a short collection of interesting features that guide novice users to particular SCUBA regions.

Publicly available KML files such as this are becoming more common as it becomes easier to translate the data into KML and display it in Google Sky. Guidance and full documentation can be found on the Google Code site.

## 5 Skyview on Google Sky

Skyview is a online service offered by NASA, which allows users to retrieve images from public catalogues of any part of the sky (McGlynn, 1996). They offer a huge range of datasets, ranging across a large collection of wavelength regimes.

In my efforts to learn more about Google Sky I created a script that grabbed images from the Skyview server and placed them into Google Sky.

### 5.1 Dynamic Colours and Resolution

One limitation of Google Sky's built-in non-optical wavelengths is that they have fixed colour tables. Unlike satellite imagery of the Earth, astronomical data has extremely variable brightness from point to point. A bright region in a nebula can cause the false-colour map of an infra-red image to be stretched in such a way as to black-out the fine detail in the region surrounding it.

Skyview creates images which stretch the dynamic range to fill the visible region, this means using the colour table to its fullest. This was the main motivation behind the creation of a script that would allow Skyview images to be viewed in Google Sky.

My script takes the field of view seen in Google Sky and sends the Right Ascension and Declination ranges to Skyview. The resulting image is then downloaded from the server and placed into Google Sky. The image is always at a fixed resolution (512x512 pixels), meaning that the script does not download too much data at once. It also uses the full colour table and as such the user can browse around an image and zoom in to a specific part for a more detailed view.

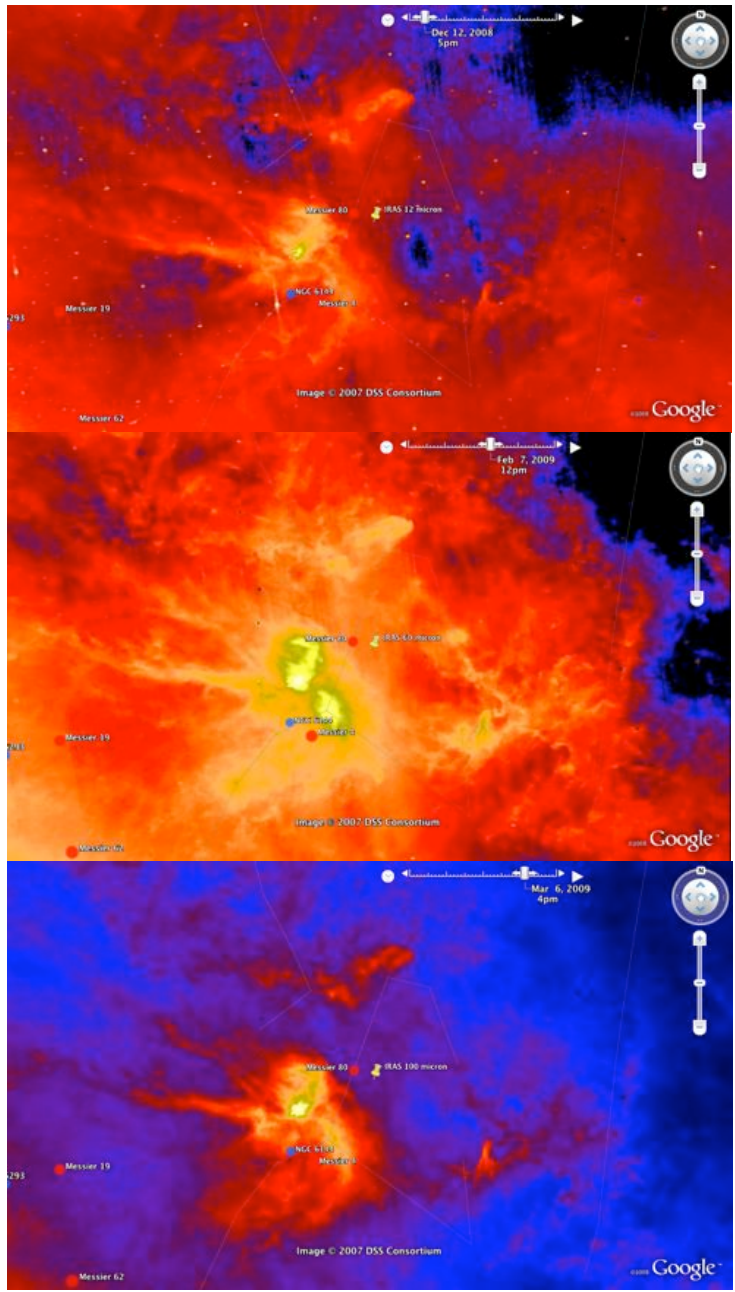

**Figure 5** – The star-forming region in Ophiuchus as seen in IRAS 12-, 60- and 100-micron bands[6] in Google Sky using the Skyview script. Different structures are picked out at different wavelengths.

## 5.2 A Wavelength Slider

Google Earth (and thus Sky) has a time slider available for datasets that change with the date or time of day. This has been used in Google Sky to show the Crab Nebula changing with time, for example.

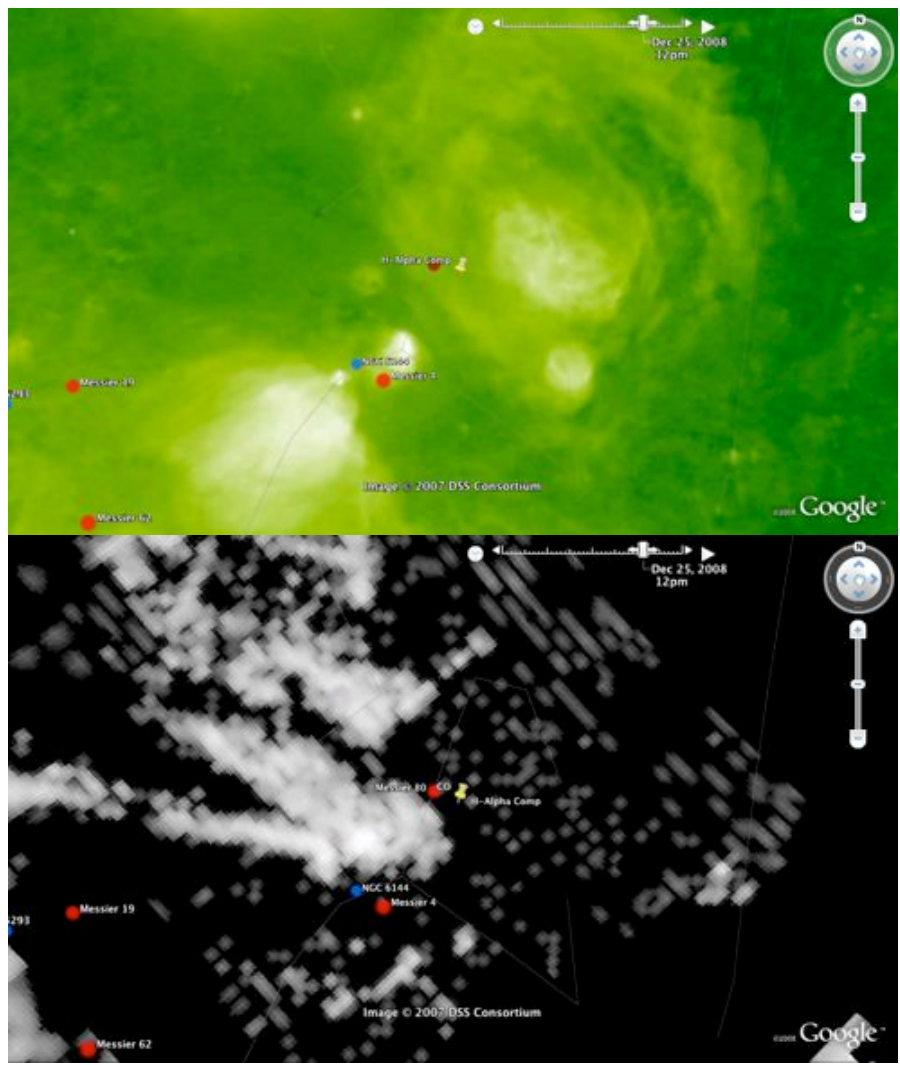

**Figure 6** – H-Alpha[3] and CO[1] images of the sky around the Ophiuchus star-forming region, just north of Scorpius. Field of view is identical to Figure 5.

However there are very few instances where time is a factor in astronomical data. I have hijacked this feature, in conjunction with the Skyview script, to create a wavelength slider in Google Sky.

Each month on the slider gets assigned a different wavelength and so if you find an interesting region it becomes possible to explore it at different frequencies. This is a feature I would very much like to see Google incorporate into the Google Earth application. It is available in a limited fashion on the Google Sky web application – but it is not yet a very sophisticated feature.

**5.3 Spectral Data on Google Sky**

Perhaps a more useful implementation of a wavelength slider in Google Sky is the idea of displaying spectral data smoothly and intuitively.
The common astronomical graphical tool, GAIA allows the user to explore data cubes (where frequency is the third dimension). In an effort to make Google Sky more feasible as a science tool, I have adapted my Skyview wavelength slider to allow the presentation of data from the HARP instrument on the JCMT.

This has been done only as a case study but it is trivial to allow full, spectral maps from HARP to be displayed in Google Sky and to slide across the spectra using the time slider.

## 6 Catching the Space Station Using Twitter

In Stuart Lowe's 'AstroTwitter' talk you can read more about how the teams working with satellites and spacecraft to relay information to the public are using Twitter. Here I would like to mention one final scripted creation, which uses Twitter in a different manner.

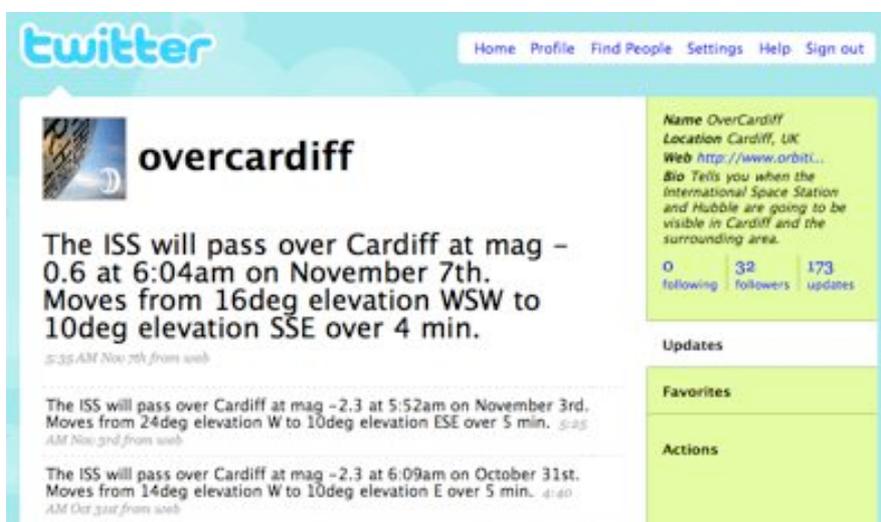

**Figure 7** – Screenshot of the OverCardiff Twitter page.

I have created Twitter accounts for many cities across the world, which Tweet when there is a visible pass of the International Space Station (ISS). Spotting the ISS is something I am often asked about my children when doing outreach work.

There is an online service called 'Heavens Above', which allows people to discover when visible transits are occurring in their area. Many find this difficult to use, or intimidating.

The Twitter feeds I created (named OverCardiff, OverDublin, OverNewYork, etc[10]) alert users to visible passes 30 minutes before they will occur. The Perl script operating this service also checks with Yahoo! weather for the current visibility in the area before sending information to Twitter. If weather conditions indicated that it is not a clear night (rain or fog, for example) then no alert is sent, since nothing will be seen.

---

[10] http://Twitter.com/overcardiff, http://Twitter.com/overdublin and
http://Twitter.com/overnewyork or for an overview, visit
http://orbitingfrog.com/blog/over-twitter/

## 7 Conclusions

There is a lot that can be done with a little knowledge of Perl or PHP. The Internet, when seen as a collection of information services, is a powerful new tool that has yet to be fully utilized.

Hopefully this article has demonstrated the type of thing that can be achieved with a little time and understanding of the technologies involved. Nothing that has been outlined here involved any advanced techniques. In fact, much of this was done to assist in my own learning of the Perl scripting language.

Google Sky and programs like it will benefit from greater use if they remain open and accessible by those who wish to create code for them. There is a large and growing community of programmers and hackers out there. Letting them contribute will create new and interesting services for the future.

## References

1. Dame, T. M., Hartmann, Dap, Thaddeus, P., ApJ, 2001, "The Milky Way in Molecular Clouds: A New Complete CO Survey", **547**, 792
2. Di Francesco, J., Johnstone, D., Kirk, H., MacKenzie, T., & Ledwosinska, E., 2008, ApJS, **175**, 277
3. Finkbeiner D.P., "A Full-Sky H-alpha Template for Microwave Foreground Prediction", 2003, ApJS, **146**, 407
4. McGlynn, T., Scollick, K., & White, N. 1996, in IAU Symp. **179**, "New Horizons from Multi-Wavelength Sky Surveys", ed. B. J. McLean et al. ( Dordrecht: Kluwer), 465
5. Scranton, R., Connolly, A., Krughoff, S., Brewer, J., Conti, A., Christian, C., McLean, B., Sosin, C., Coombe, G., Heckbert, P. 2007. "Sky in Google Earth: The Next Frontier in Astronomical Data Discovery and Visualization", arXiv e-prints, arXiv:0709.0752
6. Wheelock, S. L., et al. 1994, NASA STI/Recon Technical Report N, **95**, 22539